\documentclass[12pt, draftclsnofoot,onecolumn]{IEEEtran}   
\usepackage{amsmath}
\usepackage{amssymb}

\usepackage{graphicx}
\usepackage{epstopdf}
\usepackage{amsmath}
\usepackage{array}
\newcommand{\PreserveBackslash}[1]{\let\temp=\\#1\let\\=\temp}
\newcolumntype{C}[1]{>{\PreserveBackslash\centering}p{#1}}
\newcolumntype{R}[1]{>{\PreserveBackslash\raggedleft}p{#1}}
\newcolumntype{L}[1]{>{\PreserveBackslash\raggedright}p{#1}}
\usepackage[usenames]{color}
\usepackage{colortbl,booktabs}
\usepackage{tabularx}
\usepackage{multicol}
\usepackage{booktabs}
\usepackage[Symbol]{upgreek}
\usepackage{bm}
\usepackage{cite}
\usepackage{balance}
\usepackage[boxed]{algorithm2e}
\usepackage{mathrsfs}
\usepackage{url}
\usepackage{subfigure}
\usepackage{threeparttable}
\usepackage{amsmath}
\usepackage{dcolumn}
\usepackage{multirow}
\usepackage{booktabs}
\usepackage{amsfonts}
\usepackage{amsthm,amsmath,amssymb}
\usepackage{mathrsfs}
\usepackage{algorithmic}

\begin{document}
	\title{Wideband Channel Estimation for IRS-Aided Systems in the face of Beam Squint\vspace{-3 mm}}
	%
	\author{Siqi Ma,~\IEEEmembership{}
		Wenqian Shen,~\IEEEmembership{}
        Jianping An,~\IEEEmembership{}
        Lajos Hanzo,~\IEEEmembership{Fellow, IEEE}
		\thanks{

S. Ma, W. Shen, J. An are with the School of Information and Electronics, Beijing Institute of Technology, Beijing 100081, China (E-mails: sqma@bit.edu.com; wshen@bit.edu.cn; an@bit.edu.cn). L. Hanzo is with the Department of Electronics and Computer Science, University of Southampton, Southampton SO17 1BJ, UK (e-mail: lh@ecs.soton.ac.uk).

}
		
		\vspace{-15mm}
	}
	\maketitle
\begin{abstract}
Intelligent reflecting surfaces (IRSs) improve both the spectral and energy efficiency of wideband communication systems by using low-cost passive elements for reflecting the impinging signals with adjustable phase shifts. To fulfill the potentials of IRS-aided communication systems, accurate
channel state information (CSI) is indispensable, but it is challenging to acquire, since these passive devices cannot carry out transmit/receive signal processing. The existing channel estimation methods conceived for wideband IRS-aided communication systems only consider the channel's frequency selectivity, but ignore the effect of beam squint, despite its severe performance degradation. Hence we fill this gap and conceive the wideband channel estimation for IRS-aided communication systems by taking the effect of beam squint into consideration. We demonstrate that the mutual correlation function between the spatial steering vectors and the two-hop channel reflected by the IRS has two peaks, which leads to two angles estimated for a single propagation path, due to the effect of beam squint. One of the two estimated angles is frequency-independent `actual angle', while the other one is the frequency-dependent `false angle'. To reduce the influence of false angles on channel estimation, we propose a twice-stage orthogonal matching pursuit (TS-OMP) algorithm, where the path angles of the two-hop channel reflected by the IRS are obtained in the first stage, while the propagation gains and delays are obtained in the second stage. Moreover, we propose the corresponding bespoke pilot design by exploiting the specific the characteristics of the mutual correlation function and cross-entropy theory, for achieving an improved channel estimation performance. Our simulation results demonstrate the superiority of the proposed channel estimation algorithm and pilot design, compared to their conventional counterparts.
	\end{abstract}
	
	\IEEEpeerreviewmaketitle

\section{Introduction}\label{S1}
\IEEEPARstart{}{}
Recently, intelligent reflecting surface (IRS) aided communication systems have emerged as a promising solution for next generation systems \cite{1,2,3}, which are capable of achieving improved spectral and energy efficiency at a low cost.
In contrast to traditional amplify-and-forward (AF) relaying, IRS-aided systems use low-cost passive elements for reflecting the incident signals with adjustable phase shifts \cite{4}. By appropriately designing the phase shifts of IRS elements and the two-hop channel reflected through IRS, we can improve the communications between transceivers \cite{irs7,irs8}. There are numerous studies about beamforming design, including active beamforming at the base station (BS) and passive beamforming at the IRS for the narrow-band \cite{5,6,irs9,irs10,irs11}. Specifically, Wu {\it et al.}\cite{5} proposed a semidefinite relaxation (SDR) based method for maximizing the spectral efficiency, while Huang {\it et al.}\cite{6} advocated a gradient descent approach and sequential fractional programming method for maximizing the energy efficiency. As a further development, Feng {\it et al.} \cite{irs10} conceived a deep reinforcement learning based framework for solving the non-convex optimization problem of passive beamforming design at the IRS. Moreover, a geometric mean decomposition-based method \cite{irs12}, a convex optimization-based technique \cite{irs13} and block coordinate descent iterative algorithms \cite{irs14} were proposed for the joint passive and active beamforming design in wideband IRS-aided systems. All these contributions relied on the idealized simplifying assumption that the channel state information (CSI) is perfectly known. However, in reality channel estimation for IRS-aided communication systems is very challenging. That is because the large number of reflective elements are passive devices, which cannot perform active transmit/receive signal processing. We fill this gap in the literature by conceiving an efficient channel estimation method for the IRS-aided communication systems.

Although, there is paucity of related solutions, some researchers have designed channel estimation for IRS-aided narrow-band communication systems \cite{7,8,9,11}. Specifically, Mishra {\it et al.} \cite{7} proposed an `on-off' state control based channel estimation method, where only a single element of reflecting surface was switched on, while all other elements remained off at each time slot. In this way, the channel reflected through the activated element can be estimated without interference from the signals reflected by all other elements of the IRS. Wang {\it et al.} \cite{8} proposed a three-phase channel estimation method for the uplink of IRS-aided multiuser systems, which exploited the fact that the IRS elements reflect the signals arriving from different users to the BS via the same IRS to BS channel. Lin \cite{9} {\it et al.} proposed a Lagrange optimation based channel estimation strategy for minimizing the mean-squared error of channel estimation.
However, the above methods require a large number of measurements for distinguishing the large number of reflective elements at the IRS. To circulate this problem, Wang \cite{11} {\it et al.} formulated the channel estimation of IRS-aided systems as a sparse signal recovery problem by exploiting the channel sparsity in the angular domain. Then, they applied popular compressed sensing (CS) algorithms, such as the basis pursuit and simultaneous orthogonal matching pursuit techniques for recovering the channel at a much reduced number of measurements.

To expound a little further, some authors have also extended channel estimation from the narrow-band regime to wideband scenarios\cite{12,13,14,irs15}. Specifically, Zheng {\it et al.} \cite{12,13} formulated a wideband channel estimation problem for the IRS-aided orthogonal frequency-division multiplexing (OFDM) systems and proposed a least-square (LS) based method for estimating frequency-selective fading channels. To reduce the overhead of channel training, Yang {\it et al.}\cite{irs15} proposed an IRS elements grouping method, where each group consists of a set of adjacent IRS elements that share a common reflecting coefficient.
Wan \cite{14} {\it et al.} designed a CS-based wideband channel estimation method by assuming that the path-angle of BS to IRS channel is known. Then, they utilized a distributed orthogonal matching pursuit algorithm for estimating the IRS to user channel. However, these existing wideband channel estimation methods\cite{12,13,14,irs15} only consider the frequency-selectivity of the wideband channel, but ignore the effect of beam squint. Beam squint of wideband systems  will lead two distinct angles of two-hop channel reflected through the IRS for the same propagation path, which makes traditional channel estimation methods ineffective. To the best of our knowledge, this is the first paper studying channel estimation for IRS-aided communication systems considering the effect of beam squint. The main contributions of this paper are summarized as follows:

\begin{itemize}
\item  We formulate the estimation problem of the reflected channel, which is defined as the cascaded BS-to-IRS and IRS-to-user channel, in the face of beam squint. Specifically, the cascaded channel is characterized by the equivalent angles, gains and delays of the associated propagation paths. In contrast to the conventional channel model, where the spatial steering vectors are frequency-independent, we consider the steering vectors to be frequency-dependent. Therefore, the angular range of the steering vectors of the reflected channel is no longer equivalent to $[-1/2,1/2)$ \cite{10}. Instead, we should consider the extended angular range of steering vectors spanning $(-1,1)$.
\item We study the effect of beam squint on the channel estimation of IRS-aided communication systems and propose a twice-stage OMP (TS-OMP) algorithm, which is robust to beam squint. A popular technique is to search for the peak of correlation function between the spatial steering vectors and the reflected channel, which determines the equivalent angles. However, due to the extended angular range of steering vector, two angles will be returned for a single propagation path. One of the two estimated angles is the frequency-independent actual angle, while the other one is the frequency-dependent false angle, which may substantially erode the performance of channel estimation. To eliminate the influence of false angles, we propose a TS-OMP algorithm. Specifically, in the first stage, we exploit the fact that frequency-dependent false angles vary across the subcarriers, while the actual angles remain constant across the subcarriers. Then, we propose a block-sparse method for estimating the actual equivalent angles, while suppressing the influence of the frequency-dependent false angles. In the second stage, based on the equivalent angles estimated in the first stage, we propose an OMP based method for calculating the gains and delays of the equivalent paths having different equivalent angles.
\item  Finally, we propose the corresponding pilot design, where a pair of pilot design requirement are considered. Firstly, to eliminate the interference imposed by false angles in the first stage, we have to reduce the accumulated values corresponding to the false angles of the mutual correlation function across the different subcarriers. This means that the false angles at the pilot-subcarriers corresponding to the certain path should be separated from each other. Secondly, a high grade of orthogonality is preferred among the columns of the measurement matrix for the estimation of channel gains and delays in the second stage. Based on the above mentioned pair of requirements, we propose a cross-entropy based pilot design method, which improves the cascaded  BS-IRS-user channel estimation performance. Finally, our simulation results demonstrate that the proposed channel estimation algorithm and pilot design combination outperforms its conventional counterpart.

\end{itemize}

The rest of this paper is organized as follows. In Section II, we present the system model of IRS-aided wideband systems considering the effect of beam squint. In Section III, we analyze the effect of beam squint on estimation of BS-IRS-user channel. In Section IV, we propose a TS-OMP method for estimating the channel, which is robust to beam squint, while in Section V, we propose a pilot design based on cross-entropy. In Section VI, our simulation results are provided, followed by our conclusion in Section VII.

\emph{Notations}: We use the following notations throughout the paper. We let a, $\mathbf{a}$, $\mathbf{A}$ represent the scalar, vector, and matrix respectively; $\rm vec\{\cdot \}$ denotes the vectorization of a matrix and $\rm ivec\{\cdot \}$ denotes the invectorization of a vector;
$(\cdot)^{\rm{T}}$,
$(\cdot)^{\text{H}}$, and $(\cdot)^{-1}$ denote the transpose,
conjugate transpose, and inverse of a matrix, respectively; $\left[\mathbf{a}\right]_{m:n}$ denotes the $m$-th to $n$-th element of vector $\mathbf{a}$; $\left[\mathbf{A}\right]_{m,n}$ denotes the $(m,n)$-th element of matrix $\mathbf{A}$; $\left[\mathbf{A}\right]_{:,n}$ denotes the $n$-th column of matrix $\mathbf{A}$; The operator $ \circ $, $\otimes$ and $*$ represent the Hadamard-product, Kronecker product and convolution, respectively; Finally, $\mathbf{0}$ denotes the zero matrix, $\mathbf {I}$ denotes identity matrix, and ${\mathbf{1}}_{m,n}$ denotes all 1 matrix of size $m \times n$.

\section{System Model With Beam Squint}\label{S2}
\begin{figure}[t]
\center{\includegraphics[height=2.3in,width=3.3in]{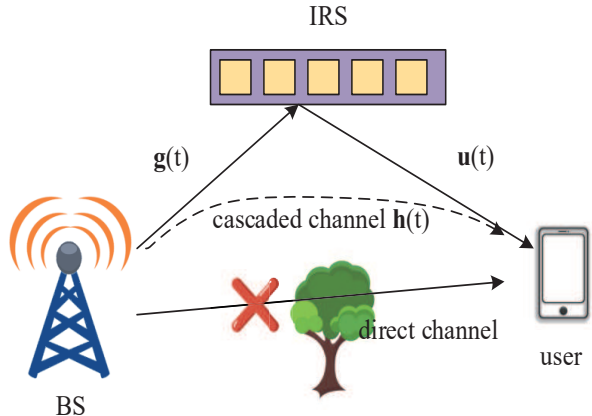}}
\caption{IRS-aided communication system}
\label{ula}
\end{figure}
	
As depicted in Fig. 1, we consider the IRS-aided wideband OFDM system, where both the BS and the user have a single antenna- \footnote{ The channel estimation method proposed in this paper can be readily extended to  multiple-antennas BS and multiple received single-antenna users by using orthogonal pilots.}, while the IRS is a ($M \times 1$)-element uniform linear array (ULA) \cite{15}. Let $\mathbf{g}(t) \in \mathbb{C}^{M \times 1}$ denote the channel impulse response (CIR) spanning from the BS to the IRS, and $\mathbf{u}(t) \in \mathbb{C}^{M \times 1}$ denote the CIR of the link from the IRS to the user. In this paper, we neglect the direct path between the BS and the user. In fact, if the direct path exists, we can estimate it by turning off the IRS and using traditional channel estimation methods \cite{11}. Based on this assumption, the direct channel can be cancelled from the IRS-aided channel model. OFDM with $N_p$ subcarriers is adopted for combating the multipath effects. We define the transmission bandwidth as $W$, yielding the subcarrier of $W/{N_p}$. We assume that the cyclic prefix (CP) is longer than the maximum multipath delay and all the complex-valued channels remain approximately constant within the channel's coherence time \cite{13}.

Assume furthermore that there are $L_1$ propagation paths between the BS and the IRS, where $\tau _{{l_1},m}^{\rm {TR}}$ denotes the time delay of the $l_1$-th path from the BS to the $m$-th IRS elements, where $l_1 \in \{1, 2,\cdots, L_1\}$ and $m \in \{1, 2,\cdots, M\}$. Thus, the CIR between the BS and the $m$-th IRS element can be expressed as \cite{irs16}
\begin{align}\label{eq_i1} 
{g_{m}}(t) = \sum\limits_{{l_1} = 1}^{{L_1}} {{{\overline \alpha  }_{{l_1}}}} {e^{ - j2\pi {f_c}\tau _{{l_1},m}^{\rm {TR}}}}\delta (t - \tau _{{l_1},m}^{\rm {TR} }),
\end{align}
where $f_c$ is the carrier frequency and ${\overline \alpha  }_{{l_1}}$ is the complex path gain of the $l_1$-th path. We define the channel vector ${\mathbf{g}}(t)=[{g_{1}}(t),\cdots,{g_{M}}(t)]^{\rm T}\in \mathbb{C}^{M \times 1}$.
Similar to the definition of ${g_{m}}(t)$, the CIR between the $m$-th IRS element and the user can be expressed as
\begin{align}\label{eq_i2} 
{u_{m}}(t) = \sum\limits_{{l_2} = 1}^{{L_2}} {{{\overline \beta  }_{l_2}}} {e^{ - j2\pi {f_c}\tau _{l2,m}^{\rm {RR}}}}\delta (t - \tau _{{l_2},m}^{ \rm {RR}}),
\end{align}
where $L_2$ and ${\overline \beta  }_{l_2}$ denote the number of paths between the IRS as well as the user and the complex gain of the $l_2$-th path, $l_2 \in \{1,\cdots,L_2\}$. We denote the delay of the $l_2$-th path from the $m$-th IRS element to the user by $\tau _{l2,m}^{\rm {RR}}$ and define the channel vector by ${\mathbf{u}}(t)=[{u_{1}}(t),\cdots,{u_{M}}(t)]^{\rm T}\in \mathbb{C}^{M \times 1}$.


\begin{figure}[ht]
\centering
\subfigure[Illustrate AOA ${\chi _{{l_1}}}$.]{               
\includegraphics[width=7cm]{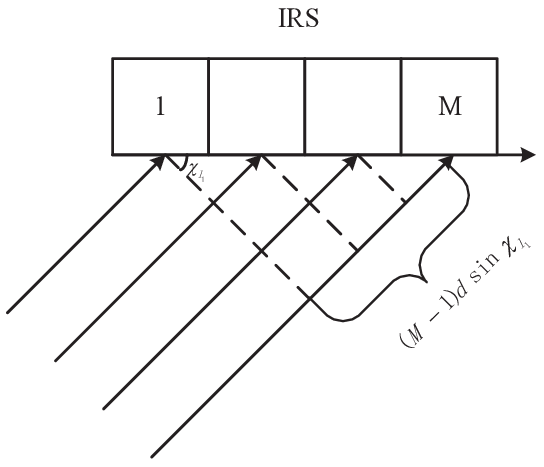}}
\hspace{0in}
\subfigure[Illustrate DOA ${\vartheta _{{l_2}}}$.]{
\includegraphics[width=7cm]{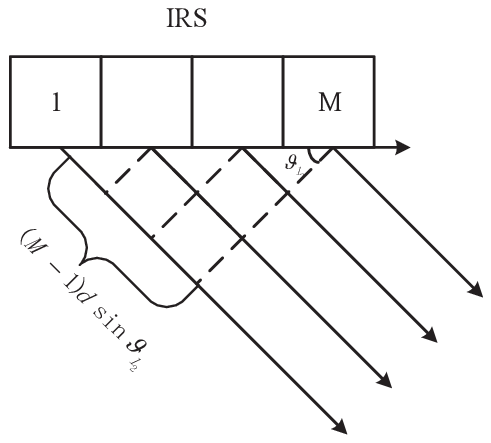}}
\caption{The illustration of AOA and DOA}
\end{figure}

Let us express the signal reflected by the $m$-th IRS element as
\begin{align}
{r_m}(t) & = {\theta _m}{g_m}(t)*s(t) \nonumber \\
\label{eq_i4} & = {\theta _m}\sum\limits_{{l_1} = 1}^{{L_1}} {{{\overline \alpha  }_{{l_1}}}{e^{ - j2\pi {f_c}\tau _{{l_{1,m}}}^{\rm{TR}}}}s(t - \tau _{{l_{1,m}}}^{\rm {TR}})},
\end{align}
where ${\pmb{\theta }} = {[{\theta _1}, \cdots ,{\theta _m}, \cdots ,{\theta _M}]^T} = {[{\varepsilon _1}{e^{j{\phi _1}}}, \cdots ,{\varepsilon _m}{e^{j{\phi _m}}}, \cdots ,{\varepsilon _M}{e^{j{\phi _M}}}]^T}$ represent the reflection coefficients of
the IRS, where ${\phi _m} \in [0,2\pi)$ and ${\varepsilon _m} \in [0,1]$ are the phase shift and amplitude reflection coefficient of the $m$-th IRS element, respectively. To maximize the signal power reflected by the IRS, we set ${\varepsilon _m}=1, {\forall _m} \in M$ \cite {5}. Moreover, $s(t)$ is the signal transmitted by the BS.

Then, the signal $r_m(t)$ is reflected by the IRS to the user and the received signal is expressed as
\begin{align}
{y_m}(t) & = {u_m}(t)*{r_m}(t)+n_m(t)\nonumber \\
\label{eq_i3} & = \sum\limits_{{l_2} = 1}^{{L_2}} {{{\overline \beta  }_{{l_2}}}{e^{ - j2\pi {f_c}\tau _{{l_2,m}}^{\rm{RR}}}}} {r_m}(t - \tau _{{l_2,m}}^{\rm{RR}})+n_m(t),
\end{align}
where $n_m(t)$ is the additive complex Gaussian noise. By substituting (\ref{eq_i4}) into (\ref{eq_i3}), we have
\begin{align}
{y_m}(t) = {\theta _m}&\sum\limits_{{l_1} = 1}^{{L_1}} {\sum\limits_{{l_2} = 1}^{{L_2}} {{{\overline \alpha  }_{{l_1}}}{{\overline \beta  }_{{l_2}}}} } {e^{ - j2\pi {f_c}\tau _{l_1,m}^{\rm{TR}}}}{e^{ - j2\pi {f_c}\tau _{{l_2,m}}^{\rm{RR}}}}\cdot s(t - \tau _{l_1,m}^{\rm{TR}} - \tau _{{l_2,m}}^{\rm{RR}})+n_m(t) \nonumber \\
\label{eq_i6} = {\theta _m}&{h_m}(t)*s(t)+n_m(t),
\end{align}
where CIR of the $m$-th cascaded BS-IRS-user element channel is written as
\begin{align}\label{eq_i7} 
{h_m}(t) =& \sum\limits_{{l_1} = 1}^{{L_1}} {\sum\limits_{{l_2} = 1}^{{L_2}} {{{\overline \alpha  }_{{l_1}}}{{\overline \beta  }_{{l_2}}}} } {e^{ - j2\pi {f_c}\tau _{{l_1},m}^{\rm{TR}}}}{e^{ - j2\pi {f_c}\tau _{{l_2},m}^{\rm{RR}}}} \cdot \delta (t - \tau _{{l_1},m}^{\rm{TR}} - \tau _{{l_2},m}^{\rm{RR}}).
\end{align}

 For the sake of simplicity, we define $\tau _{{l_1}}^{\rm TR}=\tau _{{l_1,1}}^{\rm TR}$ and $\tau _{{l_2}}^{\rm RR}=\tau _{{l_2,1}}^{\rm RR}$. Based on the assumption that IRS array size is much smaller than the distance between the BS and the IRS as well as the IRS-user distance, the path delay $\tau_{l_1,m}^{\rm TR}$ and $\tau_{l_2,m}^{\rm RR}$ can be expressed as
\begin{align}\label{eq_i8} 
\tau _{{l_1},m}^{\rm TR} = \tau _{{l_1}}^{\rm TR} + (m - 1)\frac{{d\sin {\chi _{{l_1}}}}}{c} = \tau _{{l_1}}^{\rm TR} + (m - 1)\frac{{\varphi _{{l_1}}^{\rm TR}}}{{{f_c}}},
\end{align}


\begin{align}\label{eq_i9} 
\tau _{{l_2},m}^{\rm RR} = \tau _{{l_2}}^{\rm RR} - (m - 1)\frac{{d\sin {\vartheta _{{l_2}}}}}{c} = \tau _{{l_2}}^{\rm RR} - (m - 1)\frac{{\varphi _{{l_2}}^{\rm RR}}}{{{f_c}}},
\end{align}
where ${\chi _{{l_1}}}$ denotes the angle of arrival (AOA) of the $l_1$-th path from the BS to the IRS, as shown in Fig. 2 (a). Similarly, ${\vartheta _{{l_2}}}$ denotes the angle of departure (AOD) of the $l_2$-th path from the IRS to the user, as shown in Fig. 2 (b). We define ${\varphi _{{l_1}}^{\rm TR}}=\frac{{d\sin {\chi _{{l_1}}}}}{{{\lambda _c}}}$ and ${\varphi _{{l_2}}^{\rm RR}}=\frac{{d\sin {\vartheta _{{l_2}}}}}{{{\lambda _c}}}$ as the normalized AOA and AOD, respectively, where $\lambda _c$ represents the carrier wavelength. For the typical half-wavelength element spacing of $d=\lambda _c/2$, we have ${\varphi _{{l_1}}^{\rm TR}} \in [-1/2,1/2)$, ${\varphi _{{l_2}}^{\rm RR}}\in [-1/2,1/2)$.

By substituting (\ref{eq_i8}) and (\ref{eq_i9}) into (\ref{eq_i7}), we have
\begin{align}\label{eq_i10} 
{h_m}(t) = &\sum\limits_{{l_1} = 1}^{{L_1}} {\sum\limits_{{l_2} = 1}^{{L_2}} {\underbrace {{{\overline \alpha  }_{{l_1}}}{e^{ - j2\pi {f_c}\tau _{{l_1}}^{TR}}}}_{{\alpha _{{l_1}}}}} } \underbrace {{{\overline \beta  }_{{l_2}}}{e^{ - j2\pi {f_c}\tau _{{l_2}}^{RR}}}}_{{\beta _{{l_2}}}}{e^{ - j2\pi (m - 1)\varphi _{{l_1}}^{TR}}}{e^{j2\pi (m - 1)\varphi _{{l_2}}^{RR}}}\delta (t - \tau _{{l_1},m}^{TR} - \tau _{{l_2},m}^{RR}),
\end{align}
where $\alpha_{l_1}={{{\overline \alpha  }_{{l_1}}}{e^{ - j2\pi {f_c}\tau _{{l_1}}^{TR}}}}$ and $\beta_{l_2}={{{\overline \beta  }_{{l_2}}}{e^{ - j2\pi {f_c}\tau _{{l_2}}^{RR}}}}$. By applying the continuous time Fourier transform to (\ref{eq_i10}), the frequency-domain (FD) cascaded channel corresponding to the $m$-th element of IRS can be written as
\begin{align}
&{h_m}(f)  = \int\limits_{ - \infty }^{ + \infty } {{h_m}(t){e^{ - j2\pi ft}}dt} \nonumber \\
&  = \sum\limits_{{l_1} = 1}^{{L_1}} {\sum\limits_{{l_2} = 1}^{{L_2}}  }{{{{{\alpha _{{l_1}}}}} {{\beta _{{l_2}}}}}}{e^{ - j2\pi (m - 1)\left(\varphi _{{l_1}}^{\rm TR} - \varphi _{{l_2}}^{\rm RR}\right)}}{e^{ - j2\pi f \left(\tau _{{l_1},m}^{\rm TR} + \tau _{{l_2},m}^{\rm RR}\right)}} \nonumber \\
\label{eq_i11} &  = {\sum\limits_{{l_3} = 1}^{{L_1L_2}} {c_{{l_3}}^{\rm{C}}{e^{ - j2\pi (m - 1)\varphi _{{l_3}}^{\rm{C}}\left( {1 + \frac{f}{{{f_c}}}} \right)}}{e^{ - j2\pi f\tau _{{l_3}}^{\rm{C}}}}} },
\end{align}
where the equivalent delay ${\tau _{{l_3}}^{\rm{C}}}$, angle $\varphi _{{l_3}}^{\rm C}$ and the complex gain ${c_{{l_3}}^{\rm{C}}}$ of the cascaded BS-IRS-user channel can be defined as
\begin{align}
\tau _{{l_3}}^{\rm C} &= \tau_{{l_1}}^{\rm TR} + \tau _{{l_2}}^{\rm RR}, \\
\varphi _{{l_3}}^{\rm C} &= \varphi _{{l_1}}^{\rm TR} - \varphi _{{l_2}}^{\rm RR},\\
c_{{l_3}}^{\rm{C}} &= {\alpha _{{l_1}}}{\beta _{{l_2}}},
\end{align}
where $l_3 \in \{1,2,\cdots,L_1L_2\}$ and the range of $\varphi _{{l_3}}^{\rm C}$ is $(-1,1)$. Let us discuss the range of $\varphi _{{l_3}}^{\rm C}$ in the next section in detail. Note that if we can obtain the parameters of $\tau _{{l_3}}^{\rm C},\varphi _{{l_3}}^{\rm C}$ and $ c_{{l_3}}^{\rm{C}} $, the cascaded FD channel can be recovered according to (\ref{eq_i11}).

Based on the convolution theorem and equation (\ref{eq_i6}), the FD signal received by the user through the $m$-th IRS element can be expressed as
\begin{align}\label{eq_i12}
{y_m}(f) =  {\theta _m}{h_m}(f)s(f)+n_m(f),
\end{align}
where $n_m(f)$ is the additive white Gaussian noise (AWGN). The FD signal $s(f)$ denotes the pilot symbol transmitted by the BS. Without loss of generality, we let $s(f)=1$. Thus, the FD signal received by the user can be expressed as
\begin{align}\label{eq_i13}
{y}(f) &= \sum\limits_{m = 1}^M {{y_m}(f)} \nonumber \\
&={\pmb{\theta }}^{\rm{T}}{\mathbf{h}}(f)+n(f),
\end{align}
where $n(f)\sim {CN}(0,{\delta}^2)$ and ${\delta}^2$ is the noise power. The cascaded channel vector ${\mathbf{h}}(f)=[h_1(f),h_2(f),\cdots,h_M(f)]^{\rm T}$ can be rewritten as
\begin{align}\label{eq_i14}
{\bf{h}}(f) = \sum\limits_{{l_3} = 1}^{{L_1}{L_2}} {c_{{l_3}}^{\rm{C}}} {\bf{a}}\left(\left(1 + \frac{f}{{{f_c}}}\right)\varphi _{{l_3}}^{\rm C}\right){e^{ - j2\pi f\tau _{{l_3}}^{\rm C}}},
\end{align}
where
\begin{align}
 &{\bf{a}}\left(\left(1 + \frac{f}{{{f_c}}}\right)\varphi _{{l_3}}^{\rm C}\right) = \nonumber \\
\label{eq_i15} &\left[1,{e^{ - j2\pi \left(1 + \frac{f}{{{f_c}}}\right)\varphi _{{l_3}}^{\rm C}}}, \cdots, {e^{ - j2\pi m \left(1 + \frac{f}{{{f_c}}}\right)\varphi _{{l_3}}^{\rm C}}},\cdots,{e^{ - j2\pi (M - 1) \left(1 + \frac{f}{{{f_c}}}\right)\varphi _{{l_3}}^{\rm C}}}\right]^{\rm T}
\end{align}
is the spatial steering vector. In contrast to the conventional channel model \cite{16,irs1}, the spatial steering vector is dependent on frequency and has the following property:

${\lim _{M \to \infty }}\frac{1}{M}{{\bf{a}}^{\rm H}}\left(\left(1 + \frac{f}{{{f_c}}}\right)\varphi _1\right){\bf{a}}\left(\left(1 + \frac{f}{{{f_c}}}\right)\varphi _2\right) = \delta \left(\varphi _1 - \varphi _2\right)$, which is termed as the angular orthogonality property. The proof is similar to that in \cite{irs6}.

The channel model derived in (\ref{eq_i11}) can be directly extended to the uniform planar array (UPA) scenario. For an IRS of $M_x \times M_y$ elements, we have
\begin{align}
{h_{{m_x},{m_y}}}(f) &= \sum\limits_{{l_1} = 1}^{{L_1}} {\sum\limits_{{l_2} = 1}^{{L_2}} {{\alpha _{{l_1}}}{\beta _{{l_2}}}{e^{ - j2\pi ({m_x} - 1)\left(u_{{l_1}}^{\rm TR} - u_{{l_2}}^{\rm RR}\right)\left(1 + \frac{f}{{{f_c}}}\right)}}} }  \nonumber \\
&\times {{e^{ - j2\pi ({m_y} - 1)\left(v_{{l_1}}^{\rm TR} - v_{{l_2}}^{RR}\right)\left(1 + \frac{f}{{{f_c}}}\right)}}{e^{ - j2\pi f\left(\tau _{{l_1}}^{TR} + \tau _{{l_2}}^{\rm RR}\right)}}},
\end{align}
where ${h_{{m_x},{m_y}}}(f)$ denotes the cascaded FD channel corresponding to the $(m_x,m_y)$-th IRS element. We have $u_l^{\rm x} = \frac{{d\cos (\overline u _{_l}^{\rm x})}}{{{\lambda _c}}}$ and $v_l^{\rm x} = \frac{{d\cos (\overline u _{_l}^{\rm x})\sin (\overline v _{_l}^{\rm x})}}{{{\lambda _c}}}$, where ${\overline u _{_l}^{\rm x}}$ and ${\overline v _{_l}^{\rm x}}$ denote the elevational angle and azimuth angle, respectively,  $l\in \{l_1,l_2\}$, ${\rm x}\in \{{\rm RR, TR}\}$.

\section{Effect Of Beam Squint on channel estimation of IRS-Aided systems}\label{S3}

In contrast to the conventional wideband channel model \cite{16,irs1}, where the steering vector is independent of frequency, in this paper, we consider the frequency-dependent steering vectors of (\ref{eq_i14}) for wideband IRS-aided systems. We observe from (\ref{eq_i14}) that the spatial angle $\left(1+\frac{f}{f_c}\right)\varphi _{{l_3}}^{\rm C}$ has different angular spread over different subcarriers, which is referred to as ``beam squint" \cite{irs6} \cite{add_0706}. The effect of beam squint will change the equivalent angle range $\varphi _{{l_3}}^{\rm C}$ of the cascaded channel, which will be discussed as follows.


%

According to (\ref{eq_i14}) and the angular orthogonality property of steering vectors, one can get the estimation of the equivalent angle $\varphi _{{l_3}}^{\rm C}$ by finding the peak of mutual correlation function $\Gamma^{f} ({x})$ between steering vectors and the cascaded channel, which can be expressed as
 \begin{align}
\Gamma^{f} (x) &= {{\bf{a}}^{\rm H}}\left(\left(1 + \frac{f}{{{f_c}}}\right)x\right){\bf{h}}(f)  \nonumber \\
 &= \sum\limits_{{l_3} = 1}^{{L_1}{L_2}} {\sum\limits_{m = 1}^M {{c_{{l_3}}}{e^{ - j2\pi f\tau _{{l_3}}^{\rm{C}}}}{e^{ - j2\pi (m - 1)\left(1 + \frac{f}{{{f_c}}}\right)(x - \varphi _{{l_3}}^{\rm{C}})}}} } \nonumber \\
\label{eq_irs1} &= \sum\limits_{{l_3} = 1}^{{L_1}{L_2}} {{c_{{l_3}}}{e^{ - j2\pi f\tau _{{l_3}}^{\rm{C}}}}\frac{{\sin \left(\pi M\left(1 + \frac{f}{{{f_c}}}\right)(x - \varphi _{{l_3}}^{\rm{C}})\right)}}{{\sin \left(\pi \left(1 + \frac{f}{{{f_c}}}\right)(x - \varphi _{{l_3}}^{\rm{C}})\right)}}},
\end{align}
where the search range is $x \in (-1,1)$, since $\varphi _{{l_3}}^{\rm{C}} \in (-1,1)$. We consider a single path of the channel $\mathbf{h}(f)$ for better elaborating on the effect of beam squint on the range of the equivalent angle $\varphi _{{l_3}}^{\rm{C}}$. Without loss of generality, we consider the path of $l_3=1$. Then equation of (\ref{eq_irs1}) can be simplified as

\begin{align}
\label{hanzo_add1} \Gamma^f (x) = {c_{{1}}}{e^{ - j2\pi f\tau _{{1}}^{\rm{C}}}} \frac{{\sin \left(\pi M\left(1 + \frac{f}{{{f_c}}}\right)(x - \varphi _{{1}}^{\rm{C}})\right)}}{{\sin \left(\pi \left(1 + \frac{f}{{{f_c}}}\right)(x - \varphi _{{1}}^{\rm{C}})\right)}},
\end{align}
where the range of $\varphi _{{1}}^{\rm{C}}$ is $(-1,1)$.

For the conventional wideband channel model disregarding the effect of beam squint, the term of $\frac{f}{f_c}$ is ignored in (\ref{hanzo_add1}). Hence, the steering vector is expressed as $\mathbf{a}(\varphi _{{l_3}}^{\rm{C}})=\left[1,{e^{ - j2\pi \varphi _{{l_3}}^{\rm C}}}\right.$ $\left.\cdots,{e^{ - j2\pi (M - 1)\varphi _{{l_3}}^{\rm C}}}\right]^{\rm T}$ and for the single path of $l_3=1$, $\Gamma^f (x) = {c_{{1}}}{e^{ - j2\pi f\tau _{{1}}^{\rm{C}}}} \frac{{\sin \left(\pi M \left(x - \varphi _{{1}}^{\rm{C}}\right)\right)}}{{\sin \left(\pi \left(x - \varphi _{{1}}^{\rm{C}}\right)\right)}}$. Since the value of M is usually large, the function $\frac{1}{M}\left| {\frac{{\sin (\pi M\xi )}}{{\sin (\pi \xi )}}} \right|$ only has non-zero value when $ \xi\in \mathbb{Z}$. Thus, we can estimate the equivalent angle $\varphi _{{l_3}}^{\rm{C}}$ of the cascaded channel $\mathbf{h}(f)$ by finding the peak of the function $\Gamma^{f} ({x})$. When we search for $x$ between $-1$ and $1$, the function $\Gamma^f (x)$ reaches its peak at $x = \varphi _{{1}}^{\rm{C}}$, while the other peak at $x=\varphi _{{1}}^{\rm{C}}+1
\;(\varphi _{{1}}^{\rm{C}}<0)$ or $x=\varphi _{{1}}^{\rm{C}}-1 \;(\varphi _{{1}}^{\rm{C}}>0)$. Then, we will get two estimated angles, one of which is the actual angle $\varphi _{{1}}^{\rm{C}}$ and the other one is separated from the actual angle by $1$, which is termed as false angle. The false angle is frequency-independent, and the steering vector ${\bf{a}}(\varphi _{{1}}^{\rm{C}})$ of the actual angle and steering vector ${\bf{a}}(\varphi _{{1}}^{\rm{C}}\pm 1)$ of false angle are equivalent. Thus, it is easy to understand that searching across the range of $x\in (-1,1)$ is equivalent to $x\in [-\frac{1}{2},\frac{1}{2})$, and the equivalent angle estimation is formulated as

\begin{align}
x = \left\{ \begin{array}{l}
\varphi _{{1}}^{\rm C}\quad \;\;\;\varphi _{{1}}^{\rm C} \in [ - \frac{1}{2},\frac{1}{2})\\
\varphi _{{1}}^{\rm C} - 1\;\;\varphi _{{1}}^{\rm C} \in [\frac{1}{2},1)\\
\varphi _{{1}}^{\rm C} + 1\;\;\varphi _{{1}}^{\rm C} \in ( - 1,-\frac{1}{2}],
\end{array} \right.
\end{align}
where $x\in [ - \frac{1}{2},\frac{1}{2})$. Explicitly, we only have to consider the range of equivalent angles in the steering vector as $[-\frac{1}{2},\frac{1}{2})$ instead of $(-1,1)$. Similar conclusions can be found in \cite{10}.

However, when we consider the effect of beam squint, the range of equivalent angles in the steering vector ${\mathbf{a}}\left(\left(1 + \frac{f}{{{f_c}}}\right)x\right)$ can no longer be equivalent to $[-\frac{1}{2},\frac{1}{2})$. That is because the false angles become frequency-dependent. Specifically, when we search for $x$ between $-1$ and $1$, we will get the actual peak at $x=\varphi _{{1}}^{\rm{C}}$ and the false peak at $x=\varphi _{{1}}^{\rm{C}}+\frac{f_c}{f+f_c}\; (-1<\varphi _{{1}}^{\rm{C}}\le 0)$ or $x=\varphi _{{1}}^{\rm{C}}-\frac{f_c}{f+f_c}\; (0<\varphi _{{1}}^{\rm{C}}<1)$, which is frequency-dependent. Thus, we will get two estimated angles, one of which is the actual angle of  $\varphi _{{1}}^{\rm{C}}$ and the other is the false angle of $\varphi _{{1}}^{\rm{C}}\pm \frac{f_c}{f+f_c}$. The false angle is separated from the actual angle by $\frac{f_c}{f+f_c}$. Thus, the squint of false angle over all subcarriers is $\frac{f_c}{0+f_c}-\frac{f_c}{W+f_c}$, which is independent of the specific equivalent angle $\varphi _{{1}}^{\rm{C}}$.

The false angle has a grave impact on channel estimation. We will elaborate by considering an example, where there is a single path with equivalent angle of ${\varphi _{{1}}^{\rm C}}=-1/6$ in cascaded FD channel $\mathbf{h}(f)$. The angular-domain index $x$ is selected from $ - 1 $ to $ 1$, the carrier frequency is $f_c=10$GHz, the number of subcarrier is $N_p=128$ and the system bandwidth is $W=500$MHz. Thus, the subcarrier frequency is $f = \frac{n_p}{N_p}W$, where $n_p \in \{0,1,\cdots,N_p-1\}$. The function $\Gamma^f ({x})$ at subcarriers $30,60,90,120$ is shown in Fig. 3. We observe that there are two peaks over the angular range of $(-1,1)$, which correspond to the index of the actual angle and the index of false angle. The index of actual angle remains $-\frac{1}{6}$ for these four subcarriers, as well as for all other subcarriers. By contrast, the index of false angle varies from about $0.78$ to $0.82$ between subcarriers $30$ and $120$. Furthermore, we note that the peak of false angle is quite comparable to that of actual angle, which will seriously interfere with the estimation of the actual angle, since we cannot readily distinguish the actual angle and false angle by searching for the peak of mutual correlation function. This motivates us to propose an efficient estimation method of the cascaded FD channel $\mathbf{h}(f)$, which is robust to beam squint.
\begin{figure}[t]
\center{\includegraphics[height=3.8in, width=3.4in]{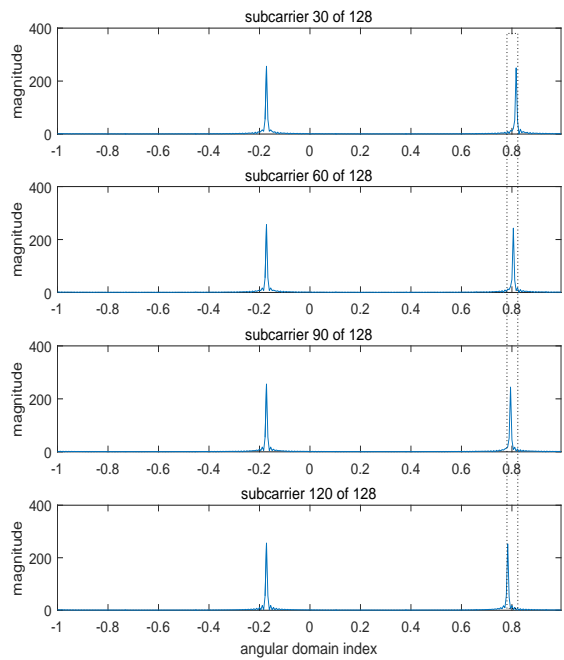}}
\caption{The mutual correlation function $\Gamma^{f} ({x})$ between steering vectors ${\bf{a}}\left((1+\frac{f}{f_c})x\right)$ and the cascaded FD channel $h(f)$ at different subcarriers, for ${\varphi _{{1}}^{\rm C}}=-1/6$ , $M=256$, $N_p=128$, $f_c=$10GHz, $W=500$MHz.}
\label{ula}
\end{figure}

\section{IRS-Aided Channel Estimation}\label{S4}

\subsection{Overview of the Proposed Channel Estimation Method}

In this section, we propose a TS-OMP based method for estimating the parameters of equivalent angles, delays and gains of cascaded channel. The philosophy of proposed method is as follows:
%
\begin{itemize}
\item In the first stage, we propose a block-sparse processing based method for estimating the equivalent angles of the cascaded channel. As shown in Fig. 3, we find that the index of the actual angle is invariant for all subcarriers, while the index of the false angle varies across all the subcarriers. Thus, if we accumulate the function $\Gamma ^{f} ({\varphi _{{1}}^{\rm C}})$ across the different subcarriers, where ${\varphi _{{1}}^{\rm C}}$ denotes the actual angle of path $l_3=1$, the accumulated value  $\Gamma ^{f} ({\varphi _{{1}}^{\rm C}})$ corresponding to the actual angel will be quite high. By contrast, when we accumulate the function $\Gamma^{f} ({\varphi _{{1}}^{\rm F}})$ across the subcarriers, where ${\varphi _{{1}}^{\rm F}}={\varphi _{{1}}^{\rm C}} \pm \frac{f_c}{f+f_c}$ denotes the false angle corresponding to the path of $l_3=1$, the value $\Gamma ^{f} ({\varphi _{{1}}^{\rm F}})$ for the angle ${\varphi _{{1}}^{\rm F}}$ accumulated over the subcarriers does not increase as fast as the accumulated value of $\Gamma ^{f} ({\varphi _{{1}}^{\rm C}})$. Therefore, we propose to accumulate the mutual correlation function $\Gamma^{f} (x)$ over the subcarriers for eliminating the interference of the false angle.
\item In the second stage, based on the estimated equivalent angles, we propose an OMP based method for estimating the equivalent path delays and gains of cascaded channel. Since there may be multiple paths having different delays for the same equivalent angle, an appropriate stopping condition is needed.
\end{itemize}

%
\subsection{Equivalent Angle Estimation for the Cascaded Channel}

Observe from (\ref{eq_i14}), that there may be multiple paths having the same equivalent angle among the $L_1L_2$ number of cascaded channel paths. Thus, (\ref{eq_i13}) can be rewritten as

\begin{align}
y(f) &= \sum\limits_{{l_3} = 1}^{{L_1}{L_2}} {{{\pmb{\theta}} ^{\rm T}}{\bf{a}}\left( {\left( {1 + \frac{f}{{{f_c}}}} \right)\varphi _{{l_3}}^{\rm C}} \right){c_{{l_3}}}{e^{ - j2\pi f\tau _{{l_3}}^{\rm {C}}}}}+n(f) \nonumber \\
\label{irs1}& = \sum\limits_{i = 1}^{{N_a}} {{\pmb{\theta}}^{\rm T}{\bf{a}}\left( {\left( {1 + \frac{f}{{{f_c}}}} \right){{\overline \varphi  }_i}} \right)\sum\limits_{{j_i} = 1}^{{J_i}} {{{\overline c }_{{j_i}}}} } {e^{ - j2\pi f{{\overline \tau  }_{{j_i}}}}}+n(f),
\end{align}
where ${\overline c _{{j_i}}}$ and ${\overline \tau  _{{j_i}}}$ denote the equivalent gain and delay of the $j_i$-th cascaded channel path, $j_i=\{1,2,\cdots, J_i\}$ and $J_i$ denotes the number of cascaded channel paths having the same equivalent angle ${{{\overline \varphi  }_i}}$, where $i=\{1,2,\cdots,N_a\}$ and $N_a$ denotes the number of different values of equivalent angles ${\varphi _{{l_3}}^{\rm C} }$ in the cascaded channel. Thus, we have $\sum_i^{N_a} J_i=L_1L_2$.

In the $n_p$-th subcarrier, i.e., $f=\frac{n_pW}{N_p}$, the signal received by the user can be expressed as
\begin{align}
\label{irs2}{{y}}({n_p}) &=  {{{\bf{{\pmb{\theta }} }}}^{\rm T}{\bf{A}}({n_p})}{\bf{z}}({n_p})+{{n}(n_p)},
\end{align}
where we denote $y(\frac{n_pW}{N_p})$ and $n(\frac{n_pW}{N_p})$ as $y(n_p)$ and $n(n_p)$ for simplicity. The sparse vector $\mathbf{z}(n_p) \in {\mathbb{C}}^{N_d \times 1}$ has $N_a$ none-zero elements. The matrix $\mathbf{A}(n_p) \in \mathbb{C}^{M\times N_d}$ is the dictionary matrix composed of $N_d$ steering vectors, which can be expressed as

\begin{align}
\label{ad_20200621}{\bf{A}}(n_p) = &\left[ {\bf{a}}\left( {\left( { - 1} \right)\left( {1 + \frac{{{n_p}W}}{{{N_p}f_c}}} \right)}\right),{\bf{a}}\left( {\left( { - 1 + \frac{2}{{{N_d}}}}\right)\left( {1 + \frac{{{n_p}W}}{{{N_p}f_c}}} \right)} \right),\right. \nonumber \\
&\cdots ,\left.{\bf{a}}\left( {\left( {1 - \frac{2}{{{N_d}}}} \right)\left( {1 + \frac{{{n_p}W}}{{{N_p}f_c}}}\right)}\right)\right] .
\end{align}

In order to estimate the equivalent angles $\{{{\overline \varphi  }_i}\}_{i=1}^{N_a}$ of the cascaded channel, we need multiple sets of observations at the user. We consider $N_s$ OFDM symbols and assume that the IRS reflection coefficients are reconfigured during different OFDM symbols. Thus, the signal received by the user in the $n_p$-th subcarrier after $N_s$ OFDM symbols can be expressed as
\begin{align}
\label{eq_i119}{\mathbf{y}}({n_p}) &=  {{{\bf{{\bf{\Theta }} }}}{\bf{A}}({n_p})}{\bf{z}}({n_p})+{\mathbf{n}(n_p)} \nonumber \\
&={{\bf{F}}({n_p})}{\bf{z}}({n_p})+{\mathbf{n}(n_p)},
\end{align}
where $\mathbf{y}(n_p)=[y^1(n_p),y^2(n_p),\cdots,y^{N_s}(n_p)]^{\rm T}$ and the element $y^{N_s}(n_p)$ denotes the signal received during the $N_s$-th OFDM symbol. The matrix ${\bf{\Theta }}=[{\pmb{\theta }}^1,{\pmb{\theta }}^2,\cdots,$ ${\pmb{\theta }}^{N_s})]^{\rm T}\in {\mathbb{C}}^{N_s\times M}$ and the vector ${\pmb{\theta }}^{N_s} \in \mathbb{C}^{M \times 1}$ denotes the reflection coefficient vector during the $N_s$-th OFDM symbol. The noise vector in (\ref{eq_i119}) is ${\mathbf{n}(n_p)}=[n^{1}(n_p),n^{2}(n_p),\cdots,n^{N_s}(n_p)]^{\rm T}$ and the element $n^{N_s}(n_p)$ represents the Gaussian noise during the $N_s$-th OFDM symbol. While we have matrix ${{\bf{F}}({n_p})}={{{\bf{{\bf{\Theta }} }}}{\bf{A}}({n_p})} \in {\mathbb{C}}^{N_s\times N_d}$. To elaborate, equation (\ref{eq_i119}) represents a sparse signal recovery problem, where $\mathbf{y}(n_p)$ is the observation vector, $\mathbf{F}(n_p)$ is the measurement matrix and $\mathbf{z}(n_p)$ is the sparse vector to be recovered. By using classic OMP algorithm, we can get the non-zero elements' indices in the vector $\mathbf{z}(n_p)$, which are expressed as ${{{\cal I}^a}}=\{ {\mathcal{I}}_i^{\rm a}\}_{i=1}^{i=N_a}$. The equivalent angles ${{{\overline \varphi  }_i}}$ can be estimated as ${{{\overline \varphi  }_i}}={ - 1 + \frac{2({\mathcal{I}}_i^{\rm a}-1)}{{{N_d}}}}$.

However, as discussed in Section III, we have to combine several subcarriers for suppressing the deleterious influence of false angles. Collecting the signal $\mathbf{y}(n_p)$ received in ${N_{{P1}}}$ subcarriers where $N_{P1}$ is the number of pilot subcarriers, we have
\begin{align}
\label{eq_i17} \overline {\bf{y}} = \overline{\bf{F}} \overline {\bf{z}}  + \overline{\bf{n}},
\end{align}
where
\begin{align}
\overline {\bf{F}}  = \left[ {\begin{array}{*{20}{c}}
{{\bf{F}}({{n}}_1 )}&{\bf{0}}& \cdots &{\bf{0}}\\
{\bf{0}}&{{\bf{F}}({{n}}_2 )}& \cdots &{\bf{0}}\\
 \vdots & \vdots & \ddots & \vdots \\
{\bf{0}}&{\bf{0}}& \cdots &{{\bf{F}}({{n}}_{N_{P1}} )}
\end{array}} \right] \in {{\mathbb{C}}^{{N_s}{N_{{P1}}} \times {N_{{P1}}}{N_d}}},
\end{align}
and $\overline {\bf{y}}=[{\bf{y}}( {{n}}_1 )^{\rm T},{\bf{y}}( {{n}}_2 )^{\rm T},\cdots,{\bf{y}}( {{n}}_{N_{P1}} )^{\rm T}]^{\rm T}\in \mathbb{C}^{N_sN_{P1}\times 1}$, $\overline {\bf{z}}=[{\bf{z}}( {{n}}_1 )^{\rm T},{\bf{z}}( {{n}}_2 )^{\rm T},\cdots,{\bf{z}}( {{n}}_{N_{P1}} )^{\rm T}]^{\rm T}\in \mathbb{C}^{N_{P1}N_a\times 1}$, $\overline {\bf{n}}=[{\bf{n}}( {{n}}_1 )^{\rm T},{\bf{n}}( {{n}}_2 )^{\rm T},\cdots,{\bf{n}}( {{n}}_{N_{P1}} )^{\rm T}]^{\rm T}\in \mathbb{C}^{N_sN_{P1}\times 1}$. Equation (\ref{eq_i17}) also represents a sparse signal recovery problem. Since the actual angle is the same for all subcarriers, the vector $\overline {\bf{z}}$ exhibits inherent block sparsity. By exploiting the block-sparsity, we propose a TS-OMP algorithm for eliminating the effect of false angles and get an accurate estimate of the cascaded channel angles, which is summarized in Algorithm 1.

In Step 1 of Algorithm 1, we apply an elementary transformation to (\ref{eq_i17}) for locating those particular elements of vector $\overline {\bf{z}}$ that represent the same equivalent angle in different subcarriers, which are then lumped together.. Specifically, this operation can be expressed as $[\widetilde {\bf{z}}]_{(n_d - 1){N_{{P1}}} + {{n}}_p} = [\overline{\bf{z}}]_{( {{n}}_p  - 1){N_d} + n_d}$, where $n_p=\{1,2,\cdots,N_{P1}\}$  and $n_d=\{1,2,\cdots,N_d\}$. To ensure that equation (\ref{eq_i17}) holds, we apply a similar transformation to the columns of matrix $\overline{\bf{F}}$, which can be expressed as $[\widetilde {\bf{F}}]_{:,(n_d - 1){N_{{P1}}} + {{n}}_p} =[ \overline {\bf{F}}]_{:,( {{n}}_p  - 1){N_d} + n_d}$. Thus, (\ref{eq_i17}) can be rewritten as $\overline {\bf{y}}  = \widetilde {\bf{F}}\widetilde {\bf{z}} + \overline {\bf{n}}$. In this way, the $N_{P1}$ adjacent elements of vector $\widetilde {\bf{z}}$ are either all zero elements or all non-zero elements. In Step 3, we estimate the indices of non-zero elements of vector $\widetilde {\bf{z}}$. Traditionally, we can obtain the indices by finding the maximum value of objective function (OF), which can be expressed as ${\cal I}_i^a = \mathop {\arg \max }\limits_{{\cal I}_i^a \in \{ 1,2, \cdots ,{N_d}\} } \sum\limits_{k = 1}^{{N_{P1}}} {\left\| {{{({{[{{\widetilde {\bf{F}}}_{{b_1}}}]}_{:,t}})}^{\rm{H}}}{\bf{r}}_a^{i - 1}} \right\|_2^2}$, where $t=({\cal I}_i^a-1)N_{P1}+k$ and the expression ${\bf{r}}_a^{i - 1}$ can be found in Step 8, while ${{\widetilde {\bf{F}}}_{{b_1}}}$ denotes the buff matrix processing in Step 6, where we set ${{\widetilde {\bf{F}}}_{{b_1}}}={{\widetilde {\bf{F}}}}$ in Step 1 of Algorithm 1. We find that there are many zeros elements in each column of matrix ${\widetilde {\bf{F}}}$. To reduce the complexity, we remove those items of ${\bf{r}}_a^{i - 1}$ that interact with the zero elements of ${\bf{F}}_{{b_1}}$. The OF can be rewritten as ${\cal I}_i^a = \mathop {\arg \max }\limits_{{\cal I}_i^a \in \{ 1,2, \cdots ,{N_d}\} } \sum\limits_{k = 1}^{{N_{P1}}} {\left\| {{{({{[{{\widetilde {\bf{F}}}_{{b_1}}}]}_{j:q,t}})}^{\rm{H}}}{{[{\bf{r}}_a^{i - 1}]}_{j:q}}} \right\|_2^2}$, where $j=(k-1)N_s+1$, $q=kN_s$ and $t=({\cal I}_i^a-1)N_{\rm P1}+k$. Repeat Steps 3 to 9 of Algorithm 1 until the stop criterion is met, when we can get the index set ${{{\cal I}^a}}=\{{\cal I}^a\}_{i=1}^{N_d}$.

Then, we can use least squares (LS) estimator to estimate the non-zero elements of vector $\mathbf{z}(n_p)$, which can be expressed as

\begin{align} \label{eq_a5_1}
{\rm LS}:\quad {[{\bf{z}}( { {{n}} }_p )]_{{{\cal I}^a}}} = {\left( {{{[{\bf{F}}( { {{n}} }_p )]}_{:,{{\cal I}^a}}}} \right)^\dag }{\bf{y}}( { {{n}} }_p ),
\end{align}
where
\begin{align}
{({[{\bf{F}}({ {{n}} }_p )]_{:,{{\cal I}^a}}})^\dag } = {\left[{({[{\bf{F}}( { {{n}} }_p )]_{:,{{\cal I}^a}}})^{\rm{H}}}{[{\bf{F}}( { {{n}} }_p )]_{:,{{\cal I}^a}}}\right]^{ - 1}}{({[{\bf{F}}({ {{n}} }_p )]_{:,{{\cal I}^a}}})^{\rm{H}}}
\end{align}
denotes the pseudo-inverse of ${[{\bf{F}}( {n} _p )]_{:,{{\cal I}^a}}}$.

\subsection{Delay- and Gain- Estimation of the Cascaded Channel}
According to (\ref{irs1}) and (\ref{irs2}), the $i$-th non-zero element of vector $\mathbf{z}(n_p)$ can be expressed as
\begin{align}
{[{\bf{z}}({n_p})]_{{{\cal I}^a_i}}} = \sum\limits_{{j_i} = 1}^{{J_i}} {{{\overline c }_{{j_i}}}{e^{ - j2\pi \frac{{{n_p}W}}{{{N_P}}}{{\overline \tau  }_{{j_i}}}}}}.
\end{align}
Upon collecting ${[{\bf{z}}({n_p})]_{{{\cal I}^a_i}}}$ of pilot-subcarriers, we have
\begin{align}
\label{irs3} {{\bf{z}}^{{{\cal I}^a_i}}} = \sum\limits_{{j_i} = 1}^{{J_i}} {{{\overline c }_{{j_i}}}\overline{\bf{b}} } \left( {{{\overline \tau  }_{{j_i}}}} \right),
\end{align}
where ${{\bf{z}}^{{{\cal I}^a_i}}} = {\left[ {{{[{\bf{z}}({{{n}} }_1 )]}_{{\cal I}_i^a}},{{[{\bf{z}}({ {{n}} }_2 )]}_{{\cal I}_i^a}}, \cdots ,{{[{\bf{z}}( { {{n}} }_{{N_{{P1}}}} )]}_{{\cal I}_i^a}}} \right]^{\rm{T}}}$ and $ \overline{\bf{b}} \left( {{{\overline \tau  }_{{j_i},i}}} \right)$ can be expressed as
\begin{align}
 \overline{\bf{b}} \left( {{{\overline \tau  }_{{j_i}}}} \right) = {\left[ {{e^{ - j2\pi \frac{{{n_1}W}}{{{N_P}}}{{\overline \tau  }_{{j_i}}}}},{e^{ - j2\pi \frac{{{n_2}W}}{{{N_P}}}{{\overline \tau  }_{{j_i}}}}}, \cdots ,{e^{ - j2\pi \frac{{{n_{{N_{P1}}}}W}}{{{N_P}}}{{\overline \tau  }_{{j_i}}}}}} \right]^{{{\rm T}}}}.
\end{align}

Equation (\ref{irs3}) can also be formulated as a sparse signal recovery problem, which can be expressed as
\begin{align}
\label{eq_a3}{\bf{z}}^{{\cal I}_i^a} = {\bf{B}}{{\bf{{\overline c }}}^{{\cal I}_i^a}},
\end{align}
where the sparse vector ${{\bf{{\overline c }}}^{{\cal I}_i^a}} \in {\mathbb{C}}^{N_{\tau} \times 1}$ has $J_i$ non-zero elements and $\mathbf{B}=[{\bf{b}}(0),{\bf{b}}(1), \cdots ,{\bf{b}}({N_\tau } - 1)]\in {\mathbb{C}}^{N_{P1}\times N_{\tau}}$ is the dictionary matrix for the delay domain composed of $N_{\tau}$ steering vectors. The vector $\mathbf{b}(k)$ of equation(\ref{eq_a3}) can be expressed as
\begin{align}
\label{eq_a7}{\bf{b}}(k) = {[{e^{ - j2\pi \frac{W}{{{N_p}}}\frac{{k{T_\tau }}}{{{N_\tau }}}{ {{n}}}_1 }},{e^{ - j2\pi \frac{W}{{{N_p}}}\frac{{k{T_\tau }}}{{{N_\tau }}}{ {{n}}}_2 }}, \cdots ,{e^{ - j2\pi \frac{W}{{{N_p}}}\frac{{k{T_\tau }}}{{{N_\tau }}}{ {{n}}}_{{N_{{P1}}}} }}]^{\rm{T}}},
\end{align}
where $T_{\tau}$ is the maximum channel delay in the cascaded channel. In this way, we can utilize our OMP-based method to estimate the vector ${{\bf{{\overline c }}}^{{\cal I}_i^a}}$. The non-zero elements of vector ${{\bf{{\overline c }}}^{{\cal I}_i^a}}$ correspond to the channel gain ${{{\overline c }_{{j_i}}}}$ and the index corresponds to the channel delay ${{{\overline \tau  }_{{j_i}}}}$. The exact details are shown in Algorithm 1.

According to (\ref{irs1}), the FD channel response of the ${\mathbf h}(n_p)$ of the $n_p$-th subcarrier in (\ref{eq_i14}) can be expressed as
\begin{align}
\label{irs5}{\mathbf h}(f)=\sum\limits_{i = 1}^{{N_d}} {{\bf{a}}\left( {\left( {1 + \frac{f}{{{f_c}}}} \right){{\overline \varphi  }_i}} \right)\sum\limits_{{j_i} = 1}^{{J_i}} {{{\overline c }_{{j_i}}}} } {e^{ - j2\pi f{{\overline \tau  }_{{j_i}}}}}.
\end{align}
Thus, after obtain the parameters of $\{ {{{\overline \varphi  }_i}}, {{{\overline c }_{{j_i}}}}, {{{\overline \tau  }_{{j_i}}}} \}$, where $i \in \{ 1,2, \cdots ,{N_a}\}$ and $j_i=\{1,2,\cdots,J_i\}$, the cascaded FD channel $h(f)$ of all the subcarriers can be obtained according to (\ref{irs5}).
\begin{algorithm}
\caption{The proposed TS-OMP algorithm}
$\mathbf{Input}$: Received signal $\overline {\bf{y}}$. Measurement matrix $\overline {\bf{F}}$ and dictionary matrix $\mathbf{B}$.

$\mathbf{Initialization}$:
${\bf{r}}_a^0 = \overline {\bf{y}}$, ${{\cal I}^a}=[\; ]$, ${\widetilde {\bf{F}}_{{b_2}}} = [\; ]$ and iteration counter $i=1$.

\begin{algorithmic}[1]
\STATE Perform elementary transformation: $[\widetilde {\bf{F}}]_{:,(n_a - 1){N_{{P1}}} +  {{n}}_p} =[ \overline {\bf{F}}]_{:,({{n}}_p  - 1){N_a} + n_a}$ and $[\widetilde {\bf{z}}]_{(n_a - 1){N_{{P1}}} + {{n}}_p} = [\overline {\bf{z}}]_{( {{n}}_p  - 1){N_a} + n_a}$. We set ${\widetilde {\bf{F}}_{{b_1}}} = \widetilde {\bf{F}}$.
\REPEAT
\STATE Estimate angular support: ${\cal I}_i^a = \mathop {\arg \max }\limits_{{\cal I}_i^a \in \{ 1,2, \cdots ,{N_d}\} } \sum\limits_{k = 1}^{{N_{P1}}} \left\| {{{({{[{{\widetilde {\bf{F}}}_{{b_1}}}]}_{j:q,t}})}^{\rm{H}}} [{{\bf{r}}_a^{i - 1}}]_{j:q} } \right\|_2^2$
, where $j=(k-1)N_s+1$, $q=kN_s$ and $t=({\cal I}_i^a-1)N_{\rm P1}+k$.

\STATE Update index set: ${{\cal I}^a}={{\cal I}^a} \cup \{ {\cal I}_i^a\}$.
\STATE Expand matrix:

${\widetilde {\bf{F}}_{b_2}} = {\widetilde {\bf{F}}_{b_2}} \cup {[{\widetilde {\bf{F}}_{{b_1}}}]_{:,({\cal I}_i^a - 1){N_{{P1}}} + 1:{\cal I}_i^a{N_{{P1}}}}}$.
\STATE Update matrix: ${[{\widetilde {\bf{F}}_{{b_1}}}]_{:,({\cal I}_i^a - 1){N_{{P_1}}} + 1:{\cal I}_i^a{N_{{P_1}}}}} = {\bf{0}}$.
\STATE Calculate the value $\mathbf{z}_b$:

${{\bf{z}}_b} = {\left( {{{({{\widetilde {\bf{F}}}_{b_2}})}^{\rm{H}}}{{\widetilde {\bf{F}}}_{b_2}}} \right)^{ - 1}}{({\widetilde {\bf{F}}_{b_2}})^{\rm{H}}}\overline {\bf{y}}$.

\STATE Update residual: ${\bf{r}}_a^i = \overline {\bf{y}}  - {\widetilde {\bf{F}}_{{b_2}}}{{\bf{z}}_b}$.
\STATE Set $i=i+1$.
\UNTIL{$\frac{{\left\| {{\bf{r}}_a^{i - 2} - {\bf{r}}_a^{i - 1}} \right\|_2^2}}{{\left\| {\overline {\bf{y}} } \right\|_2^2}} \le \zeta $}
\STATE Exploit LS estimator to calculate vector $\overline {\bf{z}}$ according to (\ref{eq_a5_1}).
\FOR{$1\le i \le N_d$}
\STATE $\mathbf{Initialization}$: $\mathbf{B}_{b_1}=\mathbf{B}$, $\mathbf{B}_{b_2}=[\;]$, $\mathbf{R}_{\tau}^0={\bf{z}}_{{P1}}^{{\cal I}_i^a}$, $\bf{T}_b=[\;]$ and the counter $t=1$.
\REPEAT
\STATE Estimate delay support: $k = \mathop {\arg \max }\limits_{k \in \{ 1,2, \cdots ,{N_\tau }\} } \left\| {{{({{[{{\bf{B}}_{{b_1}}}]}_{:,k}})}^{\rm{H}}}{\bf{r}}_\tau ^{t - 1}} \right\|_2^2$
\STATE Update index set: $\mathbf{T}_b= \mathbf{T}_b \cup \{k\}$.
\STATE Expand matrix: ${{\bf{B}}_{{b_2}}} = {{\bf{B}}_{{b_2}}} \cup {[{{\bf{B}}_{{b_1}}}]_{:,k}}$.
\STATE Update matrix: ${[{{\bf{B}}_{{b_1}}}]_{:,k}} = {\bf{0}}$.
\STATE Calculate the value ${{\bf{\overline c}}_{{\cal I}_i^a}}$: ${{\bf{\overline c}}_{{\cal I}_i^a}} = {\left( {{{({{\bf{B}}_{{b_2}}})}^{\rm{H}}}{{\bf{B}}_{{b_2}}}} \right)^{ - 1}}{({{\bf{B}}_{{b_2}}})^{\rm{H}}}{\bf{z}}^{{\cal I}_i^a}$.
\STATE Update residual: ${\bf{r}}_\tau ^t = {\bf{z}}^{{\cal I}_i^a} - {{\bf{B}}_{{b_2}}}{{\bf{\overline c}}_{{\cal I}_i^a}}$.
\STATE Set $t=t+1$.
\UNTIL {$\frac{{\left\| {{\bf{r}}_\tau ^{t - 2} - {\bf{r}}_\tau ^{t - 1}} \right\|_2^2}}{{\left\| {{\bf{z}}_{{P1}}^{{\cal I}_i^a}} \right\|_2^2}} \le \zeta $}
\STATE $\mathbf{Output}$ the set ${\bf{T}}_b$ and vector ${{\bf{\overline c}}_{{\cal I}_i^a}}$.
\STATE We have the channel angle ${{{\overline \varphi  }_i}}={ - 1 + \frac{2({\mathcal{I}}_i^{\rm a}-1)}{{{N_a}}}}$,
the delay ${{{\overline \tau  }_{{j_i}}}} = \{ \frac{{{{[{{\bf{T}}_b}]}_{ {{j}}_i }}{T_\tau }}}{{{N_\tau }}}| {{j}}_i  \in \{ 1,2, \cdots ,J_i \} \} $
and the channel gain
${{{\overline c }_{{j_i}}}} = {[{{\bf{\overline c}}_{{\cal I}_i^a}}]_{j_i }}$.
\ENDFOR
\end{algorithmic}
\KwOut{${{{\overline \varphi  }_i}}, {{{\overline c }_{{j_i}}}}, {{{\overline \tau  }_{{j_i}}}}$}
\end{algorithm}



\section{Pilot Design}\label{S5}
In this section, we propose a cross-entropy based pilot design for improving the estimation performance of the channel's equivalent angles, delays and gains.
The pilot design guidelines have to consider both the first stage and the second stage of the proposed TS-OMP algorithm, which are as follows:
\begin{itemize}
\item In the first stage of the TS-OMP algorithm, we estimate the equivalent angles of cascaded FD channel. We define $v_{{n_1}} = \varphi _1^{\rm{C}} \pm \frac{{{f_{{n_{1}}}}}}{{{f_{{n_1}}} + {f_c}}}$ and $v_{{n_{_{{N_{P1}}}}}} = \varphi _1^{\rm{C}} \pm \frac{{{f_{{n_{{N_{P1}}}}}}}}{{{f_{{n_{{N_{P1}}}}}} + {f_c}}}$ as the false angle corresponding to the actual angle $\varphi _1^{\rm{C}}$ at the $n_1$-th and $n_{N_{P1}}$-th subcarrier, where ${f_{n_1}} = \frac{{{n_1}W}}{{{N_P}}}$ and ${f_{{n_{{N_{P1}}}}}} = \frac{{{n_{{N_{P1}}}}W}}{{{N_P}}}$ denotes the frequency of the $n_1$-th and $n_{N_{P1}}$-th subcarrier, respectively. To eliminate the interference imposed by the false angle, we should decrease the accumulated values of function $\Gamma^{f} ({\varphi _{{l_3}}^{\rm F}})$ over the pilot- subcarriers, as mentioned in Section III. This means that the false angle $\varphi _{l_3}^{\rm F}={\varphi _{{l_3}}^{\rm C}} \pm \frac{f_c}{f+f_c}$ corresponding to a certain actual angle $\varphi _{l_3}^{\rm C}$ at different subcarriers should be at a different angular index in Fig. 3. We have defined the resolution of angular-domain dictionary matric as $\frac{2}{N_d}$ in (\ref{ad_20200621}). Thus, our first guideline is formulated as $v_{{n_1}}-v_{{n_{_{{N_{P1}}}}}}>\frac{2}{N_d}$.
\item In the second stage of our TS-OMP algorithm, we formulate the problem of estimating the equivalent channel delays and gains as a sparse signal recovery problem. Naturally, a high grade of orthogonality of the measurement matrix columns is preferred for sparse signal recovery \cite{irs3}\cite{irs4}. Hence the second guideline of pilot design should ensure that the measurement matrix $\bf{B}$ in (\ref{eq_a3}), which depends on the pilot-subcarriers, has near-orthogonal columns.
\end{itemize}

As for our first guideline of pilot design, we have
\begin{align}
\frac{{{f_c}}}{{{f_c} + {f_{n_1}}}} - \frac{{{f_c}}}{{{f_c} + {f_{{{n_{_{{N_{P1}}}}}}}}}}  \ge  \frac{2}{N_d}.
\end{align}
Thus, the span $D={f_{{n_{{N_{P1}}}}}} - {f_{{n_1}}}$ of pilots has to satisfy

\begin{align}
\label{irs10} D \ge \frac{{{f_{n_1}}({f_c} + \frac{N_d}{2}{f_c}) + f_c^2 - {f_{n_1}}(\frac{N_d}{2} - 1) - f_{n_1}^2}}{{(\frac{N_d}{2} - 1){f_c} + {f_{n_1}}}}.
\end{align}
As for our second guideline of pilot design, we have to design ${\bf{B}}$ so that ${{\bf{B}}^{\rm{H}}}{\bf{B}}$ becomes an approximately identity matrix,

\begin{align}
{{\bf{B}}^{\rm{H}}}{\bf{B}} \approx {N_{\rm P1}}{{\bf I}_{{N_\tau }}},
\end{align}
where ${{\bf I}_{{N_\tau }}}$ is the identity matrix of size $N_\tau \times N_\tau$ . Thus, our pilot design problem can formulated as
\begin{align}
\label{irs11}\mathop {\min }\limits_{\bf{B}} \mu ({\bf B}) &= \left\| {{{\bf{B}}^{\rm{H}}}{\bf{B}} - {N_{\rm P1}}{{\bf{I}}_{{N_\tau }}}} \right\|_2^2 \nonumber \\
&\rm {s.t.} \quad  (\ref{irs10}),
\end{align}
where the measurement matrix $\mathbf{B}$ is dependent of the subcarrier index of pilots $\{ {n_1},{n_2}, \cdots ,{n_{{N_{P1}}}}\}$. The optimization problem (\ref{irs11}) can then be solved by exhaustive search. However, the complexity of exhaustive search is excessive. For example, if the number of subcarriers is $128$ and the number of pilots is $6$, we need $5.4\times10^9$ searches to find the optimal pilots. Inspired-by the cross-entropy method of \cite{irs5}, we hence propose the pilot design of Algorithm 2.
\begin{algorithm}
\caption{Pilot design}
\KwIn{The number of candidates $N_c$, the number of elites $N_e$ and the number of iterations $N_{iter}$. The number of subcarriers $N_{\rm P}$ and the number of pilots $N_{\rm P1}$.}
$\mathbf{Initialization}$:
$P_{\rm B}^0=\frac{N_{\rm P1}}{N_{\rm P}}\times{\bf{1}}_{N_{\rm P},1}$
 and iteration counter $i=0$.

\For{$0\le i \le N_{iter}-1$}
{
 1. Randomly generate $N_c$ candidate vectors $\{ {\mathbf{d}_{n_c}}\} _{n_c = 1}^{N_c}$ according to ${\mathbf{P}_{\rm{B}}^{i}}$ and constraint of (\ref{irs10}), where $\mathbf{d}_{n_c} \! \in\!  \{0, 1\}^{N_{\rm{P}} \times 1}$. Generate the $N_c$ candidate measurement matrix $\{{\bf B}_{n_c}\}_{n_c=1}^{N_c}$ according to $\mathbf{d}_{n_c}$.

 2. Calculate the objective function: $ \mu ({\bf B}_{n_c}) = \left\| {{{\bf B}_{n_c}^{\rm{H}}}{\bf B}_{n_c} - {N_{\rm P1} }{{\bf{I}}_{{N_\tau }}}} \right\|_2^2$

 3. Sort the objective function $ \mu ({\bf B}_{n_c})$ in ascending order: $ \mu ({\bf B}_{d_{e,1}}) \le \mu ({\bf B}_{d_{e,2}}) \le\cdots \le\mu ({\bf B}_{d_{e,N_c}}) $;

 4. ${{\mathbf{P}_{\rm{B}}^{i + 1}}\! =\! \frac{1}{{N_e}}({\mathbf{d}_{d_{e,1}}}\! + \!{\mathbf{d}_{d_{e,2}}}\! + \! \cdots \!+ \!{\mathbf{d}_{d_{e,N_e}}})}$;

 5. $i=i+1$;

}
\KwOut{Designed pilot (the index of none-zero elements in ${\mathbf{d}_{d_{e,1}}}$).}
\end{algorithm}

In Step 1 of Algorithm 2, we define the probability vector ${\bf P}_{\rm B}^i \in \mathbb{C}^{N_{P}\times 1}$ of the $i$-th iteration, where the $n_p$-th element of ${\bf P}_{\rm B}^i$ denotes the probability that the $n_p$-th element of ${\bf{d}}_{n_c}$ is equal to $1$. The $n_p$-th element of ${\bf{d}}_{n_c} \in \mathbb{C}^{N_{\rm P}\times 1}$ indicates whether the pilot found occupies the $n_p$-th subcarrier. Specifically, if the $n_p$-th element of vector ${\bf{d}}_{n_c}$ is equal to $1$, then the $n_p$-th subcarrier is a pilot. Otherwise, if the $n_p$-th element of vector ${\bf{d}}_{n_c}$ is equal to $0$, the $n_p$-th subcarrier is not a pilot. We utilize the probability matrix ${\bf P}_{\rm B}^i \in \mathbb{C}^{N_{P}\times 1}$ to generate $N_c$ candidate vectors expressed as $\mathbf{d}_{n_c} \! \in\!  \{0, 1\}^{N_{\rm{P}} \times 1}$. Then, we check  whether the candidate pilots satisfy (\ref{irs10}). If not, we regenerate the pilots until the number of candidates reaches $N_c$. we then use the candidate pilots chosen to generate candidate measurement matrix $\{{\bf B}_{n_c}\}_{n_c=1}^{N_c}$ according to (\ref{eq_a7}). In Step 2 of Algorithm 2, we calculate the OF: $ \mu ({\bf B}_{n_c}) = \left\| {{{\bf B}_{n_c}^{\rm{H}}}{\bf B}_{n_c} - {N_{\rm P1} }{{\bf{I}}_{{N_\tau }}}} \right\|_2^2$. In Step 3 of Algorithm 2, we sort the OF $ \mu ({\bf B}_{n_c})$ in ascending order and retain the first $N_e$ elements. Then, we select the corresponding $N_e$ candidates $\{ {{\bf{B}}_{{n_c}}}\} _{{n_c} = {d_{e,1}}}^{{n_c} = {d_{e,{N_e}}}}$ as elites and record their indices as $\{{d_{e,1}},{d_{e,2}},\cdots,{d_{e,N_e}}\}$. In Step 4 of Algorithm 2, we update the probability matrix as ${{\mathbf{P}_{\rm{B}}^{i + 1}}\! =\! \frac{1}{{N_e}}({\mathbf{d}_{d_{e,1}}}\! + \!{\mathbf{d}_{d_{e,2}}}\! + \! \cdots \!+ \!{\mathbf{d}_{d_{e,N_e}}})}$. After $N_{iter}$ iterations, we arrive at the pilot subcarrier index set of $\{n_1,\cdots,n_p,\cdots,n_{N_{P1}}\}$, where $n_p$ is the index of non-zero elements in ${\mathbf{d}_{d_{e,1}}}$.


\section{Simulation Results}\label{S5}
In this section, we present our simulation results for characterizing the performance of the proposed channel estimation and pilot design method. We assume that the IRS elements of our ULA have a half-wavelength spacing. The carrier frequency is $f_c=20$ GHz, The AOA is ${\varphi _{{l_1}}^{\rm TR}}\in [-\frac{\pi}{2},\frac{\pi}{2})$ and the DOA is ${\varphi _{{l_2}}^{\rm RR}}\in [-\frac{\pi}{2},\frac{\pi}{2})$. The maximum delay spread of the cascaded channel is $200$ ns. The signal bandwidth is $W=510$ MHz and the other parameteters are set to $M=256$, $N_p=128$, $N_c=100$, $N_e=20$ and $L_1L_2=6$. We define the normalized mean-squared error (NMSE) of the cascaded channel at all subcarriers as
\begin{align}
{\rm NMSE}{_{\bf{h}}} = \frac{{E\left( {\sum\limits_{{n_p} = 1}^{{N_p}} {\left\| {\widehat {\bf{h}}(\frac{{{n_p}W}}{{{N_P}}}) - {\bf{h}}(\frac{{{n_p}W}}{{{N_P}}})} \right\|} _2^2} \right)}}{{E\left( {\sum\limits_{{n_p} = 1}^{{N_p}} {\left\| {{\bf{h}}(\frac{{{n_p}W}}{{{N_P}}})} \right\|} _2^2} \right)}},
\end{align}
where $\widehat{\bf{h}}(\frac{{{n_p}W}}{{{N_P}}})$ denotes the estimate of ${\bf{h}}(\frac{{{n_p}W}}{{{N_P}}})$.Furthermore, we define the NMSE of vector ${\bf z}(n_p)$ in (\ref{eq_i119}) and ${{\bf{{\overline c }}}^{{\cal I}_i^a}}$ in (\ref{eq_a3}) as
\begin{align}
{\rm NMSE}_{\bf{z}} = \frac{{E\left( {\sum\limits_{{n_p} = 1}^{{N_P}} {\left\| {\widehat {\bf{z}}({n_p}) - {\bf{z}}({n_p})} \right\|_2^2} } \right)}}{{E\left( {\sum\limits_{{n_p} = 1}^{{N_P}} {\left\| {{\bf{z}}({n_p})} \right\|_2^2} } \right)}},
\end{align}
\begin{align}
{\rm NMSE}_{\overline{\bf{c}}} = \frac{{E\left( {\sum\limits_{i = 1}^{{N_d}} {\left\| {{{\widehat {\bf{c}}}^{I_i^a}} - {{\overline {\bf{c}}}^{I_i^a}}} \right\|_2^2} } \right)}}{{E\left( {\sum\limits_{i = 1}^{{N_d}} {\left\| {{{\overline{\bf{c}}}^{I_i^a}}} \right\|_2^2} } \right)}},
\end{align}
where ${{\widehat {\bf{c}}}^{I_i^a}}$ and $\widehat {\bf{z}}({n_p})$ denote the estimate of ${{\overline {\bf{c}}}^{I_i^a}}$ and ${\bf{z}}({n_p})$.
\begin{figure}[t]
\center{\includegraphics[width=0.60\columnwidth]{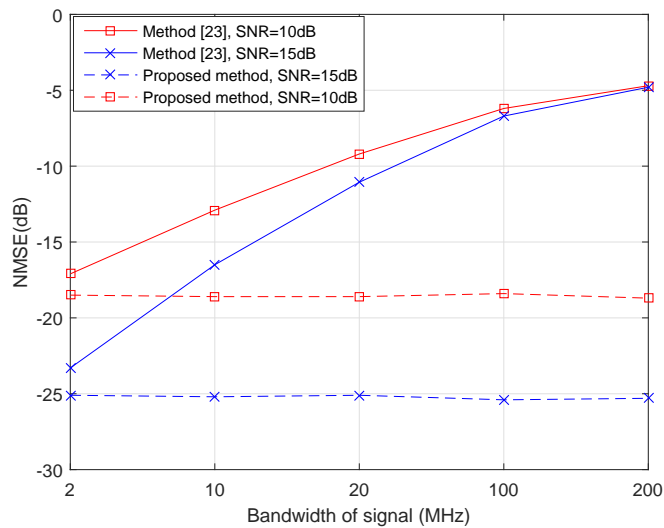}}
\caption{The NMSE performance of cascaded channel estimation against the system bandwidth. The number of IRS elements $M=256$. The number of subcarriers $N_p=128$. The frequency $f_c=20$ GHz and the pilot overhand is $10\%$.}
\label{ula}
\end{figure}

Fig. 4 shows the NMSE of the cascaded channel vs. the bandwidth. The curve {\it Method \cite{10}} denotes the NMSE of the conventional compressive sensing based method proposed in \cite{10}, which assumes that the range of equivalent angles is $[-1/2,1/2)$ and directly utilizes the OMP algorithm.
We observe that when the system bandwidth is small, the conventional method and the proposed method have similar NMSE performance. However, as the bandwidth grows, the effect of beam squint becomes severe, which degrades the performance of the traditional channel estimation method.

\begin{figure}[t]
\center{\includegraphics[width=0.58\columnwidth]{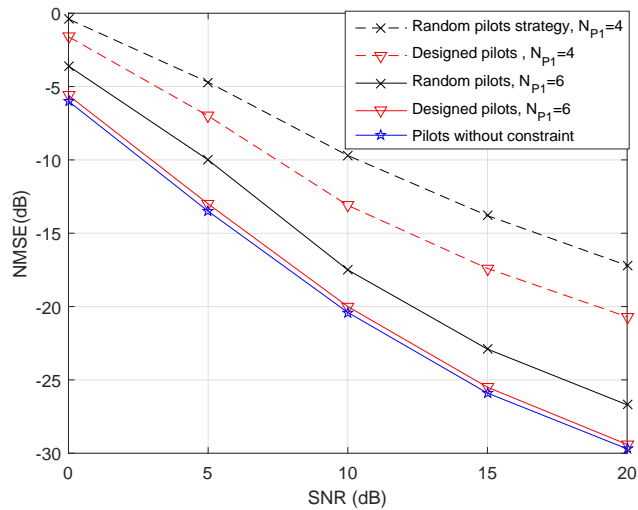}}
\caption{The NMSE vs. SNR performance of cascaded channel gains and delays estimation. The number of IRS elements $M=256$. The number of subcarriers $N_p=128$. The frequency $f_c=20$ GHz and the signal bandwidth is $510$ MHz.}
\label{ula}
\end{figure}

\begin{figure}[t]
\center{\includegraphics[width=0.58\columnwidth]{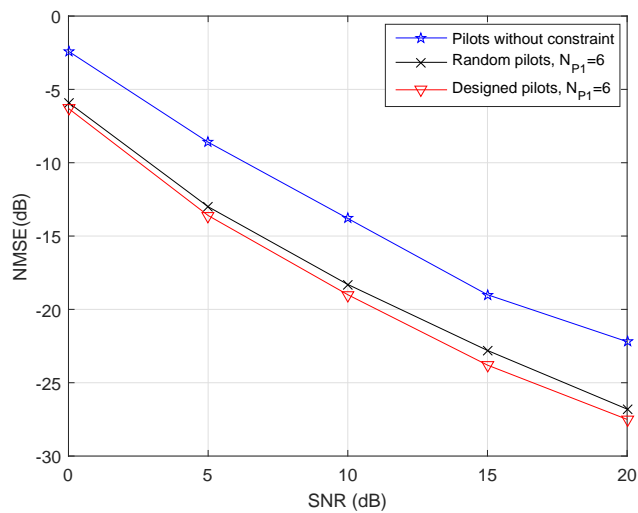}}
\caption{The NMSE vs. SNR performance of cascaded channel angles estimation. The number of IRS elements $M=256$. The number of subcarriers $N_p=128$. The frequency $f_c=20$ GHz and the signal bandwidth is $510$ MHz.}
\label{ula}
\end{figure}

Fig. 5 shows the impact of different pilot designs on the channel estimation of delays and gains vs. SNR.  The Y-axis represents the NMSE of vector ${{\bf{{\overline c }}}^{{\cal I}_i^a}}$. In Fig. 5, we estimate the channel delays and gains with known equivalent angles. The curve {\it Designed pilots, $N_{P1}=6$} and {\it Designed pilots, $N_{P1}=4$} rely on the pilots obtained by Algorithm 2, whose subcarrier indices are $\{2,20,26,43,67,91\}$ and $\{21,45,68,86\}$, respectively. The curve {\it Random pilots, $N_{P1}=6$} and {\it Random pilots, $N_{P1}=4$} correspond to the randomly selected pilots, where the number of pilots is $6$ and $4$, respectively. The curve {\it Pilots without constraint} denotes the pilots whose subcarrier indices obtained by Algorithm 2 are $\{2,4,5,6,10,12\}$, except that our first pilot design guideline is ignored. Fig. 5 shows that the proposed pilot design substantially improves the channel gain and delay estimation performance, and the constraint of (\ref{irs10}) has a minor effect on the delay and gain estimation performance of the second stage.

Fig. 6 shows the impact of different pilot designs on the equivalent angle estimation vs. SNR. The Y-axis represents the NMSE of vector ${\mathbf{z}}(n_p)$. We consider the range of equivalent angles as $(-1,1)$. Fig. 6 shows that the proposed pilot design substantially improve the algorithm's equivalent angles estimation performance. However, since the curve {\it Pilots without constraint} does not consider the first guideline, it represents poor performance. Observe in Fig. 5 and Fig. 6, that the proposed pilot design obtains both good equivalent angle as well as gain and delay estimation performance, simultaneously.


Fig. 7 shows the impact of different pilot designs on the estimation of cascaded channel against SNR. The definition of curves in Fig. 7 is the same as those in Fig. 6. Fig. 7 shows that the proposed pilot design benefically improves the algorithm's cascaded channel estimation performance.

Fig. 8 shows the NMSE performance comparison of the proposed TS-OMP and the conventional method of \cite{10} vs. the SNR. The curve {\it Proposed TS-OMP and designed pilots} represents the NMSE of the cascaded channel estimated by the proposed TS-OMP algorithm based on our designed pilots, where the range of equivalent angles is $(-1,1)$. The curve {\it Method \cite{10} and random pilots} correspond to the NMSE of the cascaded channel estimated by the conventional OMP algorithm \cite{10} based on random pilots, where the range of equivalent angles is $[-1/2,1/2)$. The curve {\it Ideal solution} represents the NMSE of channel estimation relying on known equivalent angles and delays of the cascaded channel. We observe that the proposed TS-OMP algorithm and cross-entropy based pilot design substantially improve the channel estimation performance in IRS-aided wideband systems.


\begin{figure}[t]
\center{\includegraphics[width=0.58\columnwidth]{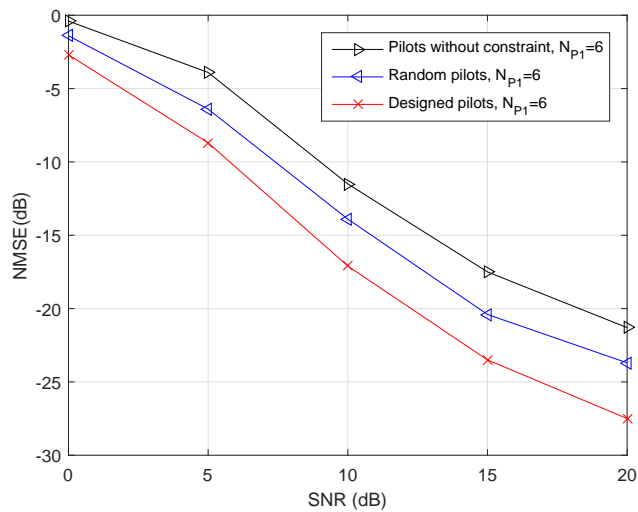}}
\caption{The NMSE vs. SNR performance of cascaded channel estimation. The number of IRS elements $M=256$. The number of subcarriers $N_p=128$. The frequency $f_c=20$ GHz and the signal bandwidth is $510$ MHz.}
\label{ula}
\end{figure}
\begin{figure}[t]
\center{\includegraphics[width=0.58\columnwidth]{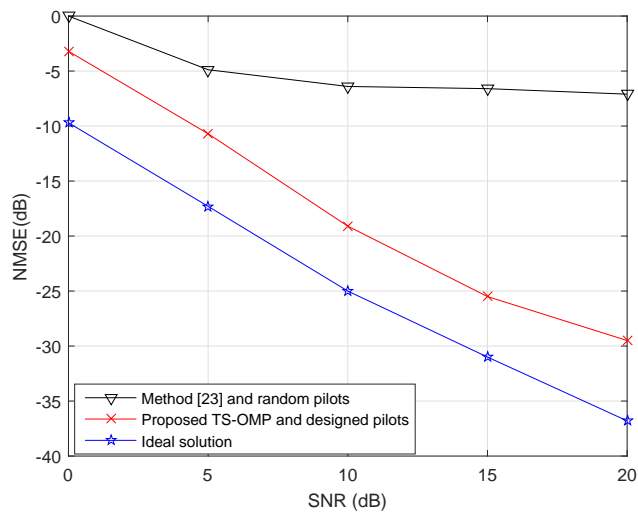}}
\caption{The NMSE vs. SNR performance of cascaded channel estimation. The number of IRS elements $M=256$. The frequency $f_c=20$ GHz and the signal bandwidth is $510$ MHz. The pilot overhand is $10\%$.}
\label{ula}
\end{figure}
Fig.9 shows the NMSE performance of the proposed channel estimation method against the number of measurements, which is defined as the number of OFDM symbols required for channel estimation. The SNR is $10$dB. The curve {\it Method \cite{10} and random pilots} shows that we cannot achieve the accurate channel estimation, not even for a high overhand of training OFDM symbols, due to the effect of beam squint. Moreover, the curve {\it Proposed method and designed pilots} indicates that the proposed method achieves good performance using a small number of training OFDM symbols.

\begin{figure}[t]
\center{\includegraphics[width=0.58\columnwidth]{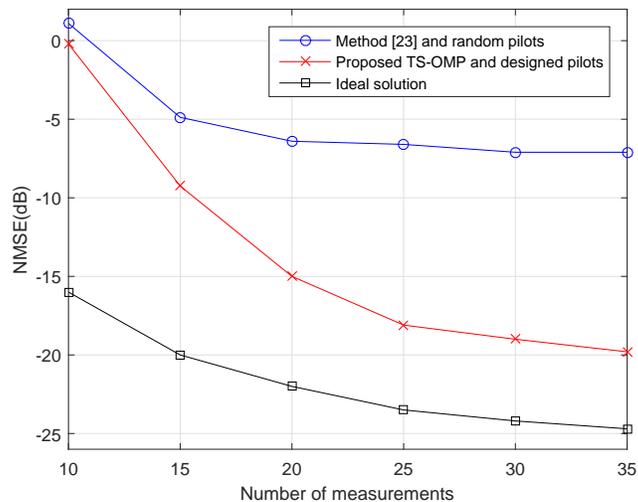}}
\caption{The NMSE performance of cascaded channel estimation against the number of measurements. The number of IRS elements $M=256$. The frequency $f_c=20$ GHz and the signal bandwidth is $510$ MHz. The SNR is 10dB. The pilot overhand is $10\%$.}
\label{ula}
\end{figure}

\begin{figure}[t]
\center{\includegraphics[width=0.58\columnwidth]{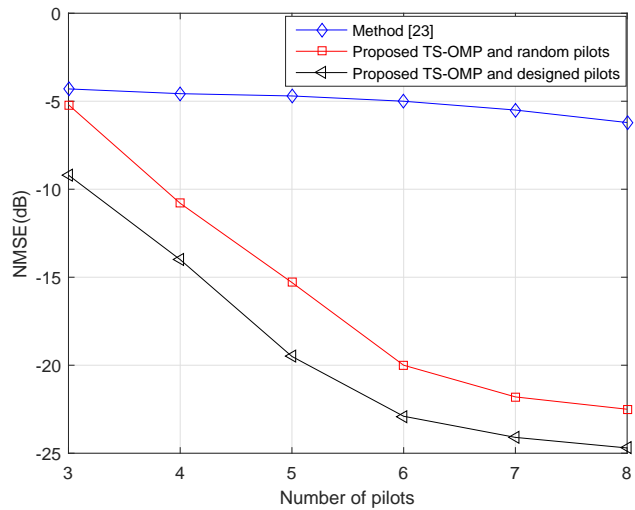}}
\caption{The NMSE performance of estimated channel against the number of pilots. The number of IRS elements $M=256$. The number of subcarriers $N_p=128$. The frequency $f_c=20$ GHz and the signal bandwidth is $510$ MHz. The SNR is 15dB.}
\label{ula}
\end{figure}
\begin{figure}[t]
\center{\includegraphics[width=0.58\columnwidth]{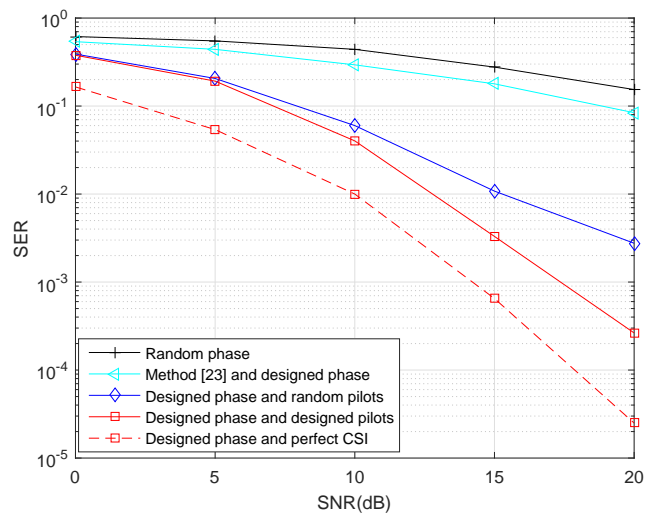}}
\caption{The SER performance against SNR. The number of IRS elements $M=256$. The number of subcarriers $N_p=128$. The frequency $f_c=20$ GHz and the signal bandwidth is $510$ MHz. The pilot overhand is $10\%$.}
\label{ula}
\end{figure}
Fig.10 shows the NMSE performance of the proposed cascaded channel estimation method against the number of pilots. The SNR is $15$dB and the number of training OFDM symbols is $35$. We find that the channel estimation becomes more and more accurate as the number of pilots increases.

Fig. 11 shows the symbol error rate (SER) against SNR. We use 16-QAM transmitted symbols.
The curve {\it Random phase} represents the SER performance, where the IRS phases are chosen randomly.
The curve {\it Method \cite{10} and designed phase} represents the SER performance when the CSI is estimated by the method proposed in \cite{10} and the phase is designed using the method of \cite{irs14}.
The curve {\it Designed phase and designed pilots} represents the SER performance, where the IRS phase is designed when the CSI is estimated by the proposed TS-OMP algorithm based on our designed pilots.
The curve {\it Designed phase and random pilots} represents the SER performance, where the IRS phase is designed when the CSI is estimated by the proposed TS-OMP algorithm based on random pilots.
The curve {\it Designed phase and perfect CSI} represents the SER performance, where the IRS phase is based on the perfect CSI.
The curve {\it Random phase} and {\it Designed phase and perfect CSI} shows that the appropriate design of the IRS phase substantially reduces the SER. The curve {\it Designed phase and perfect CSI} and {\it Designed phase and designed pilots} shows that the proposed TS-OMP algorithm based on our pilot design achieves similar SER to the perfect CSI scenario.

\section{Conclusions}\label{S5}
Cascaded channel estimation of IRS-aided systems have been investigated in wideband scenarios considering the effect of beam squint. The cascaded channel is characterized by the equivalent angles, gains and delays of propagation paths. We demonstrated that it is hard to distinguish between the actual angle and false angle, which is occurred due to the effect of beam squint. We proposed a TS-OMP algorithm, which can estimate the cascaded channel with small training overhand. Moreover, to further improve the channel estimation performance, we propose a pilot design method based on cross-entropy theory. Our simulation results confirm that the proposed TS-OMP algorithm and pilot design method achieve better performance than their conventional counterparts.
%
%

	\bibliographystyle{IEEEtran}
	\bibliography{IEEEabrv,Refference}
	
\end{document}